\documentstyle[prl,aps,multicol,epsfig]{revtex}

\begin{document}

\def \be #1{\begin{equation}\label{#1}}
\def \ee {\end{equation}}
\def \bea #1{\begin{eqnarray}\label{#1}}
\def \eea {\end{eqnarray}}
\def \Eq #1{Eq.~(\ref{#1})}
\def \Fig #1{Fig.~\ref{#1}}
\def \F2 {FPL${}^2$ }
\def \Rs {\sf I\hskip-1.5pt R} 
\def \Zs {\mbox{\sf Z\hskip-5pt Z}} 
\def \Cs {\rm C\!\!\!I\:}
\def \bp {\mbox{\boldmath $\partial$}}
\def \rb {\rm b}
\def \rg {\rm g}
\def \Kt {{\sf I\hskip-1.5pt K}}

\title{Conformational Entropy of Compact Polymers} 

\author{Jan\'{e} Kondev$^{1,2}$\footnote{janek@ias.edu} 
    and 
    Jesper Lykke Jacobsen$^{3,4}$\footnote{j.jacobsen1@physics.oxford.ac.uk}}

\address{$^1$Institute for Advanced Study, Olden Lane, Princeton, NJ 08540 \\
      $^2$Department of Physics, Princeton University, Princeton, NJ 08540 \\
         $^3$Somerville College and Department of Theoretical Physics,
             University of Oxford, 1 Keble Road, Oxford OX1 3NP, U.K. \\
         $^4$Institute of Physics and Astronomy, University of Aarhus,
             Ny Munkegade, DK-8000 Aarhus C, Denmark}

\date{\today}

\maketitle

\begin{abstract}
 Exact results for the scaling properties of compact polymers on the 
 square lattice are obtained from an effective field theory. The 
 entropic exponent $\gamma=117/112$ is calculated, and a 
 line of fixed points associated with interacting chains is
 identified;  along this line  $\gamma$  varies continuously. Theoretical 
 results are checked against detailed numerical transfer matrix calculations,
 which also yield a precise estimate for the connective constant  
 $\kappa=1.47280(1)$.  
\end{abstract}


\begin{multicols}{2}

Polymers are long,  flexible molecules, and as such their conformations 
are well described  by different types of random walks \cite{degennes_book}. 
For example, polymers in good solvents are 
modeled by self-avoiding random walks which can be mapped to a magnetic
system at the critical point. 
This mapping leads to controlled approximate calculations  for conformational 
exponents  $\nu$ and $\gamma$ in three dimensions \cite{degennes_book}, 
and exact results in two \cite{nien}. 
These exponents   describe the power-law dependence  of the linear size, 
and of the conformational entropy, on the number of monomers, respectively. 

Unlike polymers in good solvents which are swollen, globular proteins in their
native state form compact structures. On the lattice,  compact polymers are
modeled by Hamiltonian walks, i.e., self-avoiding walks that 
visit {\em all} the sites.

Recently, compact polymers  on two and three-dimensional lattices 
have become the model of choice for protein folding studies \cite{dill_rev}.
Here the focus is on the 
effect of  non-specific and non-local hydrophobic interactions among the 
amino acids, on the folding process, and on the formation of 
secondary structure (helices and sheets). 
These investigations have been almost exclusively numerical,  and   
an analytical theory of protein conformations that takes into account 
self-avoidance and compactness, as well as specific 
sequence information,  would be of considerable interest \cite{dill_rev}. 
As a first step towards this goal 
we construct a field theory of compact polymers  on the 
two-dimensional square lattice which provides detailed and exact 
information about their conformational statistics.
  
Conformational exponents for compact polymers on the 
{\em honeycomb} lattice were calculated recently using Bethe Ansatz 
techniques \cite{batch94}.
Since there is a certain degree of frustration 
associated with the Hamiltonian constraint one might expect  
the exponents  to change from one type of lattice to another. Namely, 
the number of {\em contacts}, i.e., monomer pairs  which are not adjacent 
along the chain but occupy nearest neighbor positions on the lattice, 
is {\em one} per monomer on the honeycomb, and {\em two} 
per monomer on the square lattice. 
Since hydrophobic interactions in lattice models of proteins
occur at contacts, the square and honeycomb problem describe different
physical situations.  

Here we present for the first time exact results 
for compact polymers on the square lattice.
From numerical transfer matrix calculations it was already observed
that $\gamma=1.0444(1)$ \cite{batch96} 
differs from the honeycomb value $\gamma=1$;
the latter was also the outcome of a mean-field calculation \cite{orland}.
We find  $\gamma=117/112=1.0446\ldots$. 
Furthermore, we describe a {\em line} of fixed points associated with 
interacting  compact polymers,  along which $\gamma$ varies continuously. 
These results are in excellent agreement with our numerical transfer
matrix calculations, which also provide a very precise estimate of the
connective constant, $\kappa=1.47280(1)$. This constant describes the
leading, exponential  
scaling of the number of compact polymer conformations (${\cal C}$) 
with the number of monomers: ${\cal C}\sim\kappa^{\cal N}$. 

Our identification of compact polymers with a {\em critical}
model also has bearings on the cooperativity of protein
folding thermodynamics \cite{dill_rev}. Namely, since there is no energy
gap separating the first excited (non-compact) state from
the native (compact) ones, at least in the large chain limit, 
we conclude that homopolymer collapse is a one-state process
\cite{dill_rev}.

\paragraph{Loop model}

To construct a field theory of compact polymers 
we make use of the  
two-flavor fully packed loop (FPL$^2$) model on the square lattice 
\cite{jk_prb,jk_prl}. It plays the role of the 
$O(n)$ loop model \cite{nien} from which exact results for swollen and 
dense polymers were derived. 
The polymer problem is recovered from the loop model in the 
limit of vanishing fugacity for one of the loop flavors.

The allowed loop configurations (${\cal G}$) of the \F2 model
are defined  by drawing ({\em black}) loops along the 
bonds of the square lattice with the constraint that each site is covered
by exactly one loop. Loops are not allowed to cross 
and each is assigned a fugacity $n_{\rb}$. The bonds not covered by 
loops also form ({\em gray}) loops whose number ($N_{\rg}$) 
is {\em not} constrained by the number of black loops ($N_{\rb}$). Gray loops
do not intersect, and each is  assigned a fugacity 
$n_{\rg}$. The partition function of the \F2 model  is 
\be{part1}
   Z = \sum_{\cal G} n_{\rm b}^{N_{\rm b}} n_{\rm g}^{N_{\rm g}} \ .
\ee

The \F2 model is the first loop model studied to date that possesses a
{\em two-dimensional} manifold of critical fixed points, for
$0 \le n_{\rb},n_{\rg}\le 2$ \cite{jj_npb};
$n_{\rb}\to 0$ is the compact polymer problem. 
From here on we focus on the critical region of the phase diagram, for which
exact results were previously obtained only along the line of 
special symmetry, $n_{\rb}=n_{\rg}$ \cite{jk_prl}.

\paragraph{Height model}

For the \F2 model we construct an effective field theory by mapping it to 
an interface model. The basic idea is to interpret the loops as contour
lines of a height \cite{nien}.  First we orient the 
loops independently and randomly, 
so as to be able to decide in which direction the height increases. 
Given an oriented loop configuration the microscopic heights ${\bf z}$ are
defined at the centers of the lattice plaquettes. Each bond is in one
of four states, labeled by {\em vectors} 
${\bf A}$, ${\bf B}$, ${\bf C}$ or ${\bf D}$,
depending on its flavor and direction. 
Starting from an even site an oriented black (gray) loop is
defined as a sequence ${\bf ABAB}\ldots$ (${\bf CDCD}\ldots$) of bond
states. The {\em increase} in {\bf z}, when going {\em clockwise} around an 
even site, is ${\bf A}$, ${\bf B}$, ${\bf C}$ or ${\bf D}$ depending on the 
state of the bond being crossed.
The fully packing constraint implies that all
four bond states are represented at every site, hence, in order for
the height to be well defined, we must have
${\bf A}+{\bf B}+{\bf C}+{\bf D} = 0$. The four vectors thus span a
three-dimensional vector space which  is the space of heights.
We adopt the normalization of  Ref.~\cite{jk_prb}: 
${\bf A}=(-1,+1,+1)$, ${\bf B}=(+1,+1,-1)$, ${\bf C}=(-1,-1,-1)$, and
${\bf D}=(+1,-1,+1)$
 
To complete the mapping from loops to heights we must specify the way in 
which the fugacities $n_{\rb}$ and $n_{\rg}$ are distributed between the 
two possible orientations. The clockwise oriented black  loops are assigned  
the weight $\exp({\rm i}\pi e_{\rb})$, and similarly for the 
gray loops with $e_{\rg}$ replacing $e_{\rb}$. 
The anti-clockwise loops are weighted with the 
opposite phase. This way,  summing over the two orientations, for any 
given loop 
produces the original fugacities: 
\be{fug}
n_{\rm b} = 2 \cos(\pi e_{\rm b}) \ , \ \ \ 
n_{\rm g} = 2 \cos(\pi e_{\rm g}) \ . 
\ee

The  reason for choosing to redistribute  the loop fugacity  
in this fashion is that it allows for a {\em local} definition of 
oriented loop weights, which ultimately leads to a local effective 
field theory. In particular, 
if we assign to every right turn of an oriented black loop
the weight $\lambda_{\rb}=\exp({\rm i}\pi e_{\rb}/4)$, and 
$\lambda_{\rb}^{-1}$  for a left turn,
then the whole loop will be weighted correctly, since the difference in the 
number of right and left turns for a closed loop on the square lattice 
is $\pm 4$. The rule for the gray loops is the same with  $e_{\rg}$ 
replacing $e_{\rb}$. This assigns to each vertex of the square lattice
a weight $\lambda({\bf x})$ which is the product of the local weights
associated with the oriented black and gray loop 
passing through ${\bf x}$. 

Once the height at the origin is fixed,  oriented loop configurations are 
in a one to one correspondence with height configurations which acquire their
weights. Coarse-graining of the  microscopic heights produces the
height field ${\bf h}({\bf x})$ whose  fluctuations are described
by a conformally invariant Liouville field theory. 

\paragraph{Liouville theory}

The effective field theory for the coarse grained heights
is given by the Euclidean action
\be{action_tot}
S = S_{\rm E} + S_{\rm B} + S_{\rm L} 
\ee
where each of the three terms 
\bea{action_el_sym}
S_{\rm E} & = & \frac{1}{2} \int \! d^2{\bf x}
      \ K_{\alpha\beta} \: \bp h^{\alpha}\cdot \bp h^{\beta} \label{Sel} \\ 
S_{\rm B} & = & \frac{{\rm i}}{4 \pi} \int \! d^2{\bf x} \ ({\bf e}_0
             \cdot {\bf h}) \: {\cal R} \label{Sbo} \\
S_{\rm L} & = & \int \! d^2{\bf x} \ \sum_{{\bf e}\in{\cal R}^*_w} 
\tilde{w}_{\bf e} \exp({\rm i}
   {\bf e} \cdot {\bf h}({\bf x})) \label{SLi} \ ,   
\eea
has a concrete geometrical interpretation. 

The first, {\em elastic term}, 
accounts for the entropy of oriented loop configurations.   
The symmetries of the oriented loop model impose constraints on the 
stiffness tensor \Kt: $K_{11}=K_{33}$ and $K_{12}=K_{23}=0$ \cite{jj_npb}. 
We thus find {\em three} elastic  constants which are 
not related by symmetry. This is in contrast to all previously solved
loop models which are characterized by a single coupling constant 
\cite{nien}.

The second, {\em boundary term},  
describes the coupling of the height field to the 
scalar curvature ${\cal R}$. We are only concerned with the \F2 model 
defined on flat lattices
for which ${\cal R}$ is zero everywhere except possibly at the 
boundary. For example, on the cylinder   
${\cal R}=4\pi [\delta({\infty})-\delta({-\infty})]$,  and $S_{\rm B}$ has
the effect of inserting vertex operators $\exp(\pm{\rm i}\pi{\bf e}_0\cdot
{\bf h})$ at the two far ends. These vertex operators supply
{\em winding} loops with the weight 
$\exp[{\rm i}\pi{\bf e}_0\cdot({\bf h}(\infty)-{\bf h}(-\infty))]$, 
since these are the only loops 
that contribute to the height  difference between the two far ends.  
This extra phase factor is necessary, for  
a winding loop has an equal number of left and right turns, implying
that the local vertex weights would assign it a total weight of 1, 
regardless of direction 
or flavor. For $S_{\rm B}$ to correct this, the {\em background charge}
must be
\be{bc} 
 {\bf e}_0 = - \frac{\pi}{2} (e_{\rm g} + e_{\rm b}, 0, e_{\rm g} - 
           e_{\rm b}) \ . 
\ee

The third term, the so-called {\em Liouville potential}, is the coarse-grained 
version of the microscopic weight, $\prod_{\bf x}\lambda({\bf x})$, 
assigned to  an oriented loop configuration. 
If we define $w({\bf x})=-\ln(\lambda({\bf x}))$,  
then the operator $w({\bf x})$ is invariant under 
translations in height space that form the bcc lattice ${\cal R}_w$. As
such it can be  expanded in a  Fourier series, \Eq{SLi}, where  the  
electric charges ${\bf e}$ take their values in the reciprocal lattice
${\cal R}^*_{w}$; ${\cal R}^*_{w}$ is an fcc lattice  
with a conventional cubic cell of side $2 \pi$ \cite{jj_npb}.

The three elastic constants appearing in the action completely determine
the scaling dimensions of all operators constructed from the height.
Apart from the afore-mentioned vertex operators there are also defect
operators which correspond to {\em vortex} configurations of the height.
Vertex operators are defined by the electric charge ${\bf e}$  
while defect operators are characterized by the magnetic charge ${\bf m}$, 
which is the height mismatch around the vortex core. The scaling dimension 
of a general operator with both electric and magnetic charge 
\cite{dots} is
\be{dim}
 x({\bf e}, {\bf m}) = \frac{1}{4\pi}{\bf e}\cdot[\Kt^{-1}\cdot
            ({\bf e}-2{\bf e}_0)]+\frac{1}{4\pi} {\bf m}\cdot[\Kt\cdot
            {\bf m}] \ ;  
\ee
$\Kt$ is the stiffness tensor in \Eq{Sel}. 

To calculate the three elastic constants that make up 
$\Kt$ we turn to the  {\em loop ansatz} \cite{jk_prl}, which states
that the Liouville potential is marginal, i.e., the most relevant (in the
RG sense) vertex operators
appearing in the sum in \Eq{SLi} have dimension $x({\bf e},0)=2$. There are
four such vertex operators among the twelve shortest vectors in 
${\cal R}_w^*$: $(-\pi,0,\pm\pi)$ and $(-\pi,\pm\pi,0)$. 
Using \Eq{dim} this leads to four equations for
the three elastic constants with the unique solution:
\bea{coup}
  K_{11}  & = &  \frac{\pi}{8} \ (2-e_{\rb}-e_{\rg}), \nonumber \\
  K_{13}  & = & \frac{\pi}{8} \ (e_{\rb}-e_{\rg}), \nonumber \\
  K_{22}  & = &  \frac{\pi}{2} \ 
  \frac{(1-e_{\rb})(1-e_{\rg})}{2-e_{\rb}-e_{\rg}} \ .
\eea

\paragraph{Conformational exponents}

\begin{figure}
  \begin{minipage}{8.66cm}
   \epsfxsize=8.6cm \epsfbox{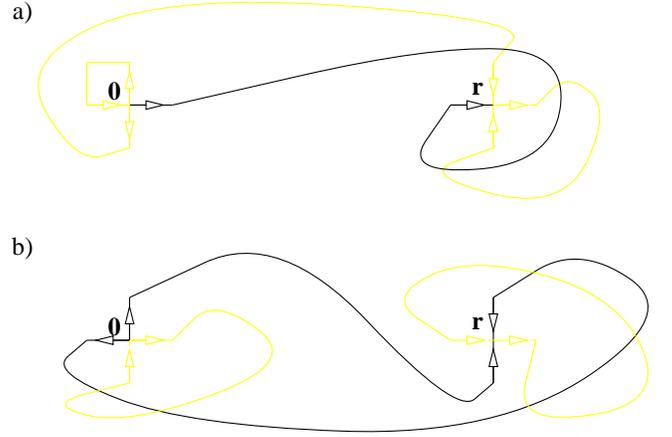}
   \caption{     
           Defect configurations used for calculating 
           geometrical scaling exponents $x_{1,1}$ (a) and $x_{2,0}$ (b).
           Bond states around point ${\bf 0}$, listed clockwise from the 
 	   leftmost bond, are ${\bf D C A C}$ (a) and ${\bf A A C D}$ (b).}
   \label{Fig:defects}
  \end{minipage}
\end{figure}

Points in the critical region of the \F2 model are 
characterized by the central charge and the geometrical scaling dimensions 
$x_{s_{\rb},s_{\rg}}$, which are both functions of the loop fugacities
$n_{\rb}$ and $n_{\rg}$. The central  charge provides information 
about the finite-size corrections to the free energy, while 
the scaling dimensions are defined by the asymptotic relation    
$Z_{s_{\rb},s_{\rg}}({\bf 0},{\bf r})/Z\sim |{\bf r}|^{-2x_{s_{\rb},s_{\rg}}}$;
$Z_{s_{\rb},s_{\rg}}({\bf 0},{\bf r})$ is the partition function of
the \F2 model with the constraint that there are $s_{\rb}$ black and 
$s_{\rg}$ gray strings connecting points ${\bf 0}$ and ${\bf r}$; 
see Fig.~\ref{Fig:defects}. We consider the two end points to be 
in the bulk, in which case $s_{\rb}+s_{\rg}$ is necessarily even; the 
odd case is associated with boundary operators \cite{jj_npb}.

The central charge of the Liouville field theory is \cite{dots}
\be{c_charge}
c = 3 + 12 x({\bf e}_0, 0) = 3 - 6\left( \frac{e_{\rm b}^2}{1 - e_{\rm b}}
      + \frac{e_{\rm g}^2}{1 - e_{\rm g}} \right) \ .
\ee
In Table~\ref{tab} we compare this formula with numerical transfer matrix
calculations along the compact polymer line ($n_{\rb}=0$, $e_{\rb}=1/2$) 
and excellent agreement is found. This holds true for the {\em whole} critical 
region \cite{jj_npb}.

Here we are interested in the conformational exponents $\gamma$ and
$\nu$ for compact polymers \cite{rem}, and so it suffices to calculate the 
geometrical scaling dimensions $x_{1,1}$ and $x_{2,0}$ \cite{saleur}.  
The formula for the  whole spectrum $x_{s_{\rb},s_{\rg}}$ is a
simple generalization of the calculation presented below \cite{jj_npb}. 

To calculate $x_{1,1}$ we consider the diagram in \Fig{Fig:defects}a 
which represents an \F2 configuration with a single black and a single 
gray string connecting the two points separated by ${\bf r}$. In the height 
representation 
these configurations are associated with topological defects with 
charges $\pm{\bf m}_{1,1}$ placed at the two points; 
${\bf m}_{1,1}={\bf C}-{\bf B}=(-2,-2,0)$ is the net height change
upon encircling the point ${\bf 0}$. Since an oriented segment is  
weighted by a complex phase, whose value depends on the number of times the  
string winds around ${\bf 0}$ and ${\bf r}$, 
vertex operators with charge  ${\bf e}_0$ have to be inserted at both 
points in order to weight all segments equally \cite{nien}. Therefore,
$x_{1,1}$ is the dimension of  an operator with total charge 
$({\bf e}_0,{\bf m}_{1,1})$, and
\bea{x1}
 2 x_{1,1} &=& \frac{1}{4} \left[ (1-e_{\rm b}) + (1-e_{\rm g}) \right] 
  \nonumber \\
       &+& \frac{(1-e_{\rm b})(1-e_{\rm g})}{(1-e_{\rm b})+(1-e_{\rm g})}
        -  \left[ \frac{e_{\rm b}^2}{1-e_{\rm b}} +
           \frac{e_{\rm g}^2}{1-e_{\rm g}} \right]  
\eea
follows from Eqs.~\ref{dim} and~\ref{coup}.

Similarly the diagram \Fig{Fig:defects}b leads to the result:
\be{x2}
 2 x_{2,0} = 2x({\bf e}_{\rm b},{\bf m}_{2,2})
       = (1-e_{\rm b}) - \frac{e_{\rm b}^2}{1-e_{\rm b}} \ .
\ee
In this case point ${\bf 0}$ serves as a source of two black strings, and
therefore corresponds to a topological defect of strength 
${\bf m}_{2,2}={\bf A}-{\bf B}=(-2,0,2)$. The electric charge 
${\bf e}_{\rm b}=-\pi/2 (e_{\rb},0,-e_{\rb})$ now compensates for the 
extraneous phase factors due to the windings of the black loop segments
only, since no gray segments connect the two points.

Our main results follow from Eqs.~(\ref{x1})--(\ref{x2})
and the scaling relations $\gamma=1-x_{1,1}$ and $1/\nu=2-x_{2,0}$ 
\cite{saleur}. For compact polymers  $\gamma=117/112$  
is obtained by
setting $e_{\rm b}=1/2$ ($n_{\rb}=0$) and $e_{\rm g}=1/3$ ($n_{\rg}=1$)   
in \Eq{x1}. The fact that $x_{1,1}<0$ indicates an effective repulsion 
between the chain ends. This is to be contrasted to 
the mean-field result $\gamma=1$, also found for compact polymers on the 
honeycomb lattice \cite{batch94}, which implies that   
the chain ends are uncorrelated. 
Hitherto numerical studies based on enumerations of short compact
polymer conformations \cite{thurmalai} 
have failed to provide the accuracy needed to
distinguish $\gamma$ from its mean-field value.

Examination of \Eq{x1} reveals a novel feature of compact
polymers on the square lattice. Namely, allowing $n_{\rg}$
to take different values along the $n_{\rb}=0$ line represents 
a situation where
different conformations are weighted differently depending
on the number of gray loops present. In the critical region of 
the \F2 model each of these weighted compact polymer problems defines a 
different critical geometry characterized by a continuously varying $\gamma$. 
Similar behavior was predicted for  
directed self-avoiding walks with orientation dependent contact interactions
\cite{cardy_gam}.

Finally, from \Eq{x2} we obtain the conformational exponent 
$\nu=1/(2-x_{2,0})=1/2$, independent of $n_{\rg}$. This result 
serves as a nice consistency check on our theory since compact structures
have Hausdorff dimension $D=2$ regardless of how they are weighted, 
and $\nu=1/D$.

\paragraph{Numerical results} 

To check our results we have constructed transfer matrices in a
connectivity basis analogous to that of the O($n$) model 
\cite{Blote89}, but taking into account the additional flavor
information of the \F2 model \cite{jj_npb}. The various sectors
containing 0, 1 and 2 strings were considered for strip widths up to
$L=14$. Conformal invariance was used to relate the finite-size
corrections of the eigenvalue spectra to the central charge and
various geometrical scaling dimensions. Results along the
compact polymer line are shown in Table~\ref{tab}; agreement with
theory is excellent, apart from discrepancies for $x_{1,1}$ at $n_{\rg}=2$ 
which we attribute to logarithmic corrections.
An extrapolation scheme based on the exact values of $c$, \Eq{c_charge},
yielded very precise estimates of the connective
constant $\kappa$. Like $\gamma$, $\kappa$ for $n_{\rg}=1$ differs
slightly from its mean-field value $4/{\rm e}=1.4715\ldots$ \cite{orland}, 
revealing the entropic origin of the effective repulsion between chain ends.
Furthermore, $\kappa$ changes continuously with the interaction--like
parameter $n_{\rg}$. 
\begin{table}
 \begin{minipage}{8.66cm}
 \begin{tabular}{c|ccccc}
  $n_{\rg}$ &  0.0        &  0.5           &  1.0
                          &  1.5           &  2.0           \\ \hline
  $c$       & -3.004(5)   & -1.815(3)      & -0.998(2)
                          & -0.411(2)      & -0.002(3)      \\
  (10)      & -3          & -1.8197$\ldots$& -1
                          & -0.4124$\ldots$&  0             \\ \hline
  $x_{1,1}$ & -0.2500(3)  & -0.1313 (9)    & -0.0439(9)
                          &  0.0255(9)     &  0.0839(5)     \\
  (11)      & -0.25       & -0.1323$\ldots$& -0.0446$\ldots$
                          &  0.0260$\ldots$&  0.1042$\ldots$\\ \hline
  $x_{2,0}$ &  0.0000(0)  &  0.0000(0)     &  0.0000(0)
                          &  0.0000(0)     &  0.0000(0)     \\
  (12)      &  0          &  0             &  0
                          &  0             &  0             \\ \hline
  $\kappa$  &  1.41422(2) &  1.44477(1)    &  1.47280(1)
                          &  1.49896(1)    &  1.52371(1)    \\
 \end{tabular}
 \caption{Numerical results for the central charge $c$, geometrical
          exponents $x_{1,1}$ and $x_{2,0}$, and the connective
          constant $\kappa$ along the line $n_{\rb}=0$. 
          Comparison is made to predictions from Eqs.~(10), (11) and (12). }
 \label{tab}
\end{minipage}
\end{table}
\nopagebreak[3]
In conclusion, we have constructed a field theory of compact polymers
on the square lattice from which exact results regarding their 
scaling properties were obtained for the first time. We hope that this
might serve as a first step towards an analytic theory of simple 
lattice models of globular proteins \cite{dill_rev}. 

Discussions with M.~Aizenman, J.~L.~Cardy, B.~Duplantier,
D.~S.~Fisher, C.~L.~Henley, T.~Hwa, T.~Spencer and C.~Zeng
are acknowledged, as is hospitality at ITP (Santa Barbara) and
grants from EPSRC (GR/J78327) and NSF (PHY94-07194 and DMS 9304580).

\end{multicols}


\begin{references}

  \bibitem{degennes_book} P.-G.~de Gennes, {\em Scaling Concepts in Polymer
                          Physics} (Cornell University Press, Ithaca, 1979).

  \bibitem{nien}          B.~Nienhuis, in {\em Phase Transitions and Critical
                          Phenomena}, edited by C.~Domb and J.~L.~Lebowitz
                          (Academic, London, 1987), Vol.~11.

  \bibitem{dill_rev}      K.~A.~Dill {\em et al.},
                          Protein Science {\bf 4}, 561 (1995).

  \bibitem{batch94}   	  M.~T.~Batchelor, J.~Suzuki and C.~M.~Yung, 
			  Phys.~Rev. Lett.~{\bf 73} 2646 (1994).


  \bibitem {batch96}      M.~T.~Batchelor, H.~W.~J.~Bl\"ote, B.~Nienhuis
                          and C.~M.~Yang, J.~Phys.~A {\bf 29}, L399 (1996).

  \bibitem{orland}       H.~Orland, C.~Itzykson and C.~de Dominicis,
                          J.~Phys.~(Paris) {\bf 46}, L353 (1985).

  
  \bibitem{jk_prb}       J.~Kondev and C.~L.~Henley,
                          Phys.~Rev.~B {\bf 52}, 6628 (1995).

  \bibitem {jk_prl}       J.~Kondev, Phys.~Rev.~Lett.~{\bf 78}, 4320 (1997).

  \bibitem {jj_npb}       J.~L.~Jacobsen and J.~Kondev, cond-mat/9804048.
                          Submitted to Nucl.~Phys.~B.

  \bibitem{dots}          Vl.~S.~Dotsenko and V.~A.~Fateev,
                          Nucl.~Phys.~B {\bf 240}, 312 (1984);
                          {\em ibid.} {\bf 251}, 691 (1985).

  \bibitem{rem}           ${\cal C}/Z\sim {\cal N}^{\gamma}$ and 
                          $R\sim{\cal N}^{\nu}$; $R$ is the radius of
                          gyration. 

  \bibitem{saleur}         B.~Duplantier and H.~Saleur,
                          Nucl.~Phys.~B {\bf 290}, 291 (1987). 

  \bibitem{thurmalai}     C.~J.~Camacho and D.~Thirumalai,
                          Phys.~Rev.~Lett.~{\bf 71}, 2505 (1993). 

  \bibitem{cardy_gam}     J.~L.~Cardy, Nucl.~Phys.~B {\bf 419}, 411 (1994). 


  \bibitem {Blote89}      H.~W.~J.~Bl\"{o}te and B.~Nienhuis,
                          J.~Phys.~A {\bf 22}, 1415 (1989).

\end{references}
\end{document}